\def\lesssim{\mathrel{\hbox{\rlap{\hbox{\lower4pt\hbox{$\sim$}}}\hbox{$<$}}}}

\def\gtrsim{\mathrel{\hbox{\rlap{\hbox{\lower4pt\hbox{$\sim$}}}\hbox{$>$}}}}

\def\msun{M$_{\odot}$}

\def\ll_lsun{log$({L/\rm L_{\odot}})$~}

\def\masa_msun{$M/ \rm M_{\odot}$~}

\def\m_mstar{$M/M_{*}$~}

\documentclass{aa}
\usepackage{graphicx}
        
\begin{document}

\title{Can pulsating PG1159 stars place constraints on the 
occurrence of core overshooting?}

\author{A. H. C\'orsico$^1$\thanks{Member of the Carrera del Investigador
Cient\'{\i}fico y Tecnol\'ogico, CONICET, Argentina.}
\and L. G. Althaus$^{1,2 \star}$}

\offprints{A. H. C\'orsico}

\institute{
$^1$   Facultad   de   Ciencias  Astron\'omicas   y   Geof\'{\i}sicas,
Universidad  Nacional de  La Plata,  Paseo del  Bosque S/N,  (1900) La
Plata,   Argentina.\\  $^2$  Departament   de  F\'\i   sica  Aplicada,
Universitat  Polit\`ecnica de  Catalunya, Av.  del Canal  Ol\'\i mpic,
s/n, 08860 Castelldefels, Spain \\
\email{acorsico,althaus@fcaglp.unlp.edu.ar} }

\date{Received; accepted}

\abstract{The present letter is aimed at exploring the influence of 
overshooting  during the  central  helium burning  in pre-white  dwarf
progenitors on  the pulsational properties  of PG1159 stars.   To this
end we follow the  complete evolution an intermediate-mass white dwarf
progenitor  from the  zero  age main  sequence  through the  thermally
pulsing and born-again phases to  the domain of the PG1159 stars.  Our
results  suggest that the  presence of  mode-trapping features  in the
period  spacings  of  these  hot  pulsating stars  could  result  from
structure in  the carbon-oxygen  core. We find  in particular  that in
order  to  get enough  core  structure  consistent with  observational
demands,  the  occurrence of  overshoot  episodes  during the  central
helium  burning is  needed.  This conclusion  is valid  for thick
helium  envelopes like  those predicted  by our  detailed evolutionary
calculations.  If the envelope  thickness were  substantially smaller,
then the  occurrence of core  overshooting would be more  difficult to
disentangle  from  the  effects  related to  the  envelope  transition
zones.
}

\authorrunning{C\'orsico \& Althaus}

\titlerunning{Pulsational constraints on the occurrence of core 
overshooting}

\maketitle

 
\section{Introduction}

Convective  overshooting is a  longstanding problem  in the  theory of
stellar  structure and  evolution.  It  is well  known  on theoretical
grounds  that  during many  stages  in  their  lives stars  experience
overshoot  episodes,  that  is  partial  mixing  beyond  the  formally
convective boundaries  as predicted by the  Schwarzschild criterium of
convective stability  (Zahn 1991; Canuto  1992; Freytag et  al.  1996;
see also Renzini 1987).  In particular, core overshooting taking place
during  central   burning  is  an  important  issue   because  it  has
significant effects  on the stellar structure and  evolution. Over the
years,   considerable  observational  effort   has  been   devoted  to
demonstrating   the   occurrence   of  core   overshooting.    Indeed,
confrontation of  stellar models with a wide  variety of observational
data suggests that convective overshoot takes place in real stars (see
Stothers \&  Chin 1992; Alongi  et al. 1993; Kozhurina-Platais  et al.
1997; Herwig et al. 1997; von Hippel \& Gilmore 2000 among others).

However,  most   of  the  evidence   about  the  occurrence   of  core
overshooting  relies primarily  on  observational data  from the  very
outer layers of stars from  where radiation emerges.  A more promising
and  direct  way of  placing  constraints  on  the physical  processes
occurring in the very deep interior  of stars is by means of the study
of  their pulsational  properties.   Pulsating white  dwarf stars  are
particularly  important  in  this   regard.   In  fact,  white  dwarfs
constitute the end product of  stellar evolution for the vast majority
of  stars,  and  the  study  of  their  oscillation  spectrum  through
asteroseismological techniques has become  a powerful tool for probing
the otherwise inaccessible inner regions of these stars (Bradley 1998;
Metcalfe et  al. 2002;  Metcalfe 2003).  White  dwarf asteroseismology
has also opened the door to peer into the physical processes that lead
to  the  formation  of   these  stars.  In  particular,  Straniero  et
al. (2003)  have raised the issue  of using pulsating  white dwarfs to
constrain the efficiency  of extra mixing episodes in  the core of the
white dwarf progenitors.

This  letter is  aimed at  specifically assessing  the  feasibility of
employing pulsating PG1159 stars to demonstrate the occurrence of core
overshoot  during the  core  helium burning  phase.  Pulsating  PG1159
stars  (or  variable  GW  Virginis) are  very  hot  hydrogen-deficient
post-AGB stars with  surface layers rich in helium,  carbon and oxygen
that exhibit $g$-mode luminosity  variations.  These stars are thought
to have experienced a very  late helium-shell flash during their early
cooling phase  after hydrogen burning has almost  ceased --- a born-again
episode; see  Fujimoto 1977, Sch\"onberner 1979. As  shown by Kawaler
\& Bradley (1994), variable PG1159 stars are particularly important to
infer fundamental  properties about pre-white dwarfs  in general, such
as  the   location  of   chemical  interfaces  and   envelope  masses.
Specifically,  we  present  an  adiabatic  pulsation  study  based  on
detailed stellar models, the evolution of which has been followed from
the  zero-age main sequence  through the  thermally pulsing  phase and
born-again episode to the PG1159  state. This fact allows us to obtain
PG1159 models  with a  physically sound internal  structure consistent
with the predictions of the theory of stellar evolution.

\section{Details of evolutionary and pulsational computations}

The adiabatic pulsational analysis presented in this work are based on
evolutionary  stellar  models  that  take into  account  the  complete
evolution  of  the progenitor  star  (see  Althaus  et al.   2005  for
details)\footnote{The stellar models  have also recently been employed
in Gautschy et al. (2005) for nonadiabatic pulsation studies of PG1159
stars.}. Specifically, the evolution  of an initially $2.7\, M_{\sun}$
stellar model  from the zero-age  main sequence has been  followed all
the way from the stages of  hydrogen and helium burning in the core up
to  the tip  of  the AGB  where  helium thermal  pulses occur.   After
experiencing 10  thermal pulses, the  progenitor departs from  the AGB
and evolves toward high  effective temperatures, where a final thermal
pulse takes  place soon after the  early white dwarf  cooling phase is
reached  --- a  very late  thermal  pulse and  the ensuing  born-again
episode, see Bl\"ocker (2001) for  a review. During this episode, most
of  the residual  hydrogen envelope  is engulfed  by  the helium-flash
convection  zone and  completely  burnt.  After  the  occurrence of  a
double-loop    in   the    Hertzsprung-Russell   diagram,    the   now
hydrogen-deficient,    quiescent   helium-burning   0.5895-$M_{\odot}$
remnant evolves at  constant luminosity to the domain  of PG1159 stars
with a surface chemical composition rich in helium, carbon and oxygen:
($^{4}$He,  $^{12}$C, $^{16}$O)=  (0.306, 0.376,  0.228).  This  is in
good agreement  with surface abundance patterns  observed in pulsating
PG1159 stars (Dreizler \& Heber 1998; Werner 2001).  Also, the surface
nitrogen abundance (about 0.01 by  mass) predicted by our models is in
line with  that detected  in pulsating PG1159  stars (see  Dreizler \&
Heber  1998).  We  mention that  abundances changes  are  described by
means of  a time-dependent  scheme that simultaneously  treats nuclear
evolution and mixing processes  due to convection and overshooting.  A
treatment of this kind  is particularly necessary during the extremely
short-lived phase of the  born-again episode, for which the assumption
of instantaneous  mixing is inadequate. Overshooting is  treated as an
exponentially  decaying  diffusion  process  and has  been  considered
during all evolutionary phases.  In particular, overshooting occurring
toward the end of core  helium burning phases yields a sharp variation
of  the chemical  composition  in the  carbon/oxygen core.   Radiative
opacities  are  those  of  OPAL  (including  carbon-  and  oxygen-rich
compositions,   Iglesias  \&  Rogers   1996),  complemented,   at  low
temperatures, with the molecular  opacities from Alexander \& Ferguson
(1994).

\begin{figure}
\centering
\includegraphics[clip,width=250pt]{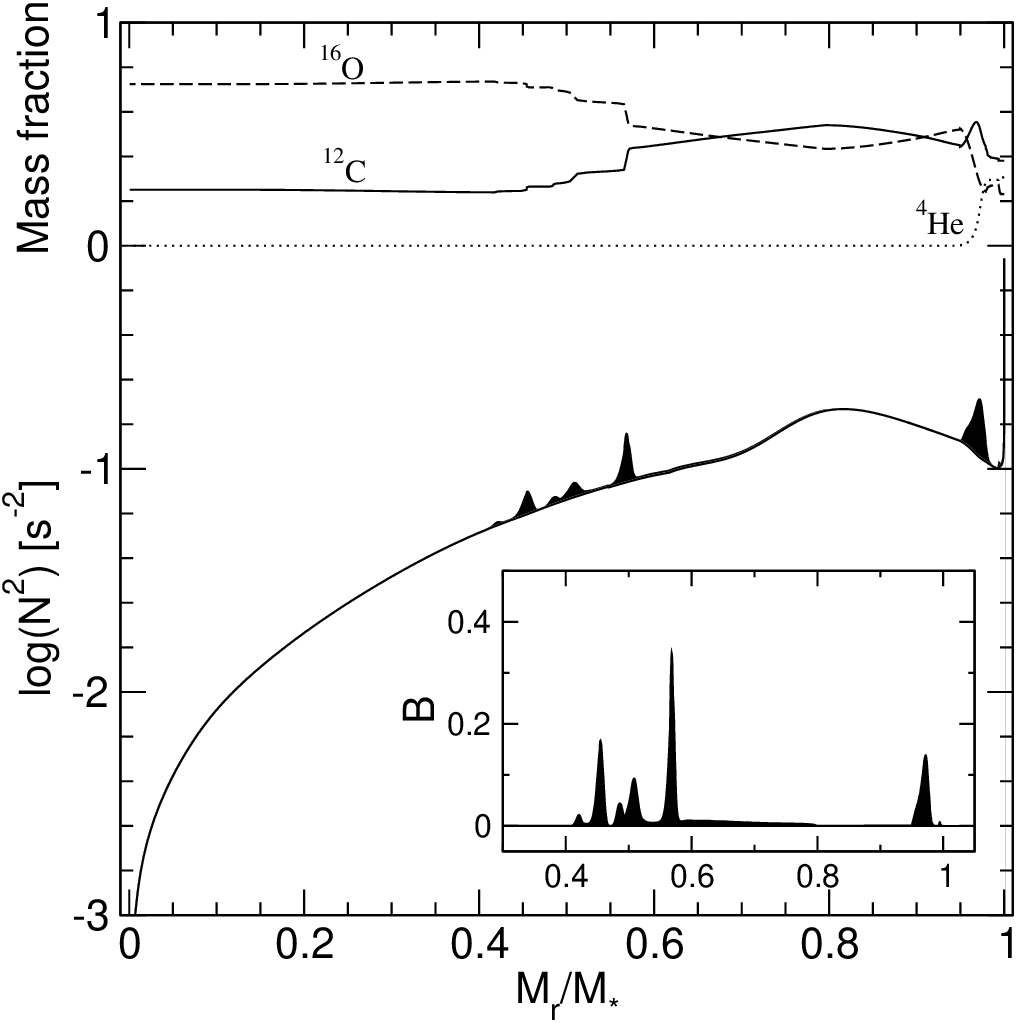}
\caption{The run of the squared Brunt-V\"ais\"al\"a frequency in terms 
of the  mass coordinate, corresponding to  a 0.5895-$M_{\odot}$ PG1159
model at  $T_{\rm eff}= 139000$ K and  $\log(L/L_{\odot})= 2.31$.  The
mass of  the helium  content is of  $0.0052 M_{\odot}$.   Dark regions
denote the contributions  of the Ledoux term $B$  (shown in the inset)
to  the Brunt-V\"ais\"al\"a  frequency.  The  chemical profile  of the
main nuclear species is displayed in the upper zone of the plot.}
\label{profile}
\end{figure}

As for pulsational calculations,  we have computed adiabatic $g$-mode
periods by employing  an updated  version of the  pulsation code  
described in C\'orsico et al.  (2001).  We limited our calculations  
to the degrees
$\ell= 1,  2$ because the  periods observed in pulsating  PG1159 stars
have been  identified with $\ell= 1,2$.   To get values  of periods as
precise as possible, we have employed about $2700-3000$ mesh-points to
describe our background stellar  models. The prescription we follow to
assess  the run  of  the Brunt-V\"ais\"al\"a  frequency  ($N$) is  the
so-called ``Ledoux  Modified'' treatment  (see Tassoul et  al.  1990),
appropriately  generalized to  include the  effects of  having several
nuclear species  with varying  abundance in a  given region.   In this
numerical  treatment  the  contribution  to  $N$ from  any  change  in
composition  is almost completely  contained in  the Ledoux  term $B$;
this fact renders the method particularly useful to infer the relative
weight that each chemical  transition region have on the mode-trapping
properties of the model.

\section{Results and discussion}

\begin{figure}
\centering
\includegraphics[clip,width=250pt]{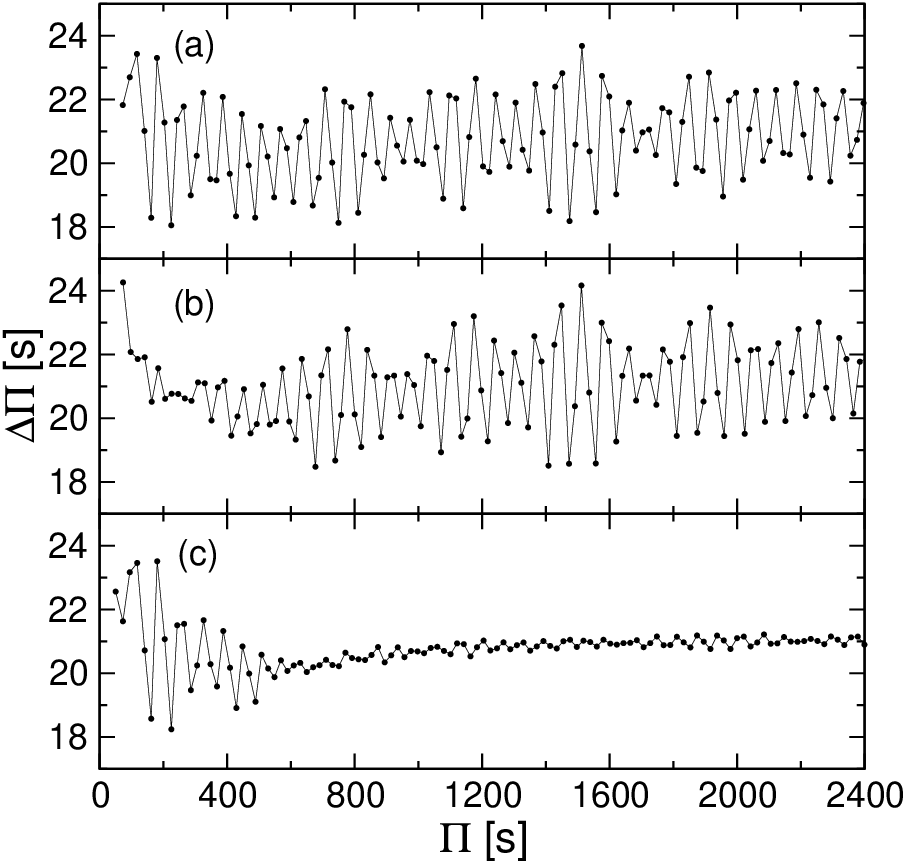}
\caption{The forward period spacing $\Delta \Pi$ vs period 
$\Pi$ for dipole ($\ell= 1$)  modes for the same PG1159 model analyzed
in  Fig.    \ref{profile}.   Panel  (a)   corresponds  to  pulsational
calculations   in   which   the    Ledoux   term   $B$   is   computed
self-consistently, whereas  panel (b) corresponds to  the situation in
which the Ledoux  term has been artificially suppressed  in the region
at $M_r/M_*  \approx 0.96$,  and panel (c)  correspond to the  case in
which $B= 0$ at $M_r/M_* \approx 0.4-0.6$.  See text for details.}
\label{deltap}
\end{figure}

We  begin by examining  Fig. \ref{profile}  in which  a representative
spatial run of the Brunt-V\"ais\"al\"a  frequency of a PG1159 model is
displayed.  The model is characterized by a stellar mass of $0.5895$
\msun,  an effective  temperature of  $\approx 139000$  K and  a 
luminosity of  $\log(L/L_{\odot})= 2.31$. In addition,  the plot shows
the internal chemical stratification of the model for the main nuclear
species (upper region of the  plot), and for illustrative purposes the
profile of  the Ledoux  term $B$ (inset).   The figure  emphasizes the
role   of   the   chemical    interfaces   on   the   shape   of   the
Brunt-V\"ais\"al\"a frequency.   At the core region  there are several
peaks at $M_r/M_* \approx  0.4-0.6$ resulting from steep variations in
the inner oxygen/carbon profile.  The  stepped shape of the carbon and
oxygen  abundance   distribution  within  the  core   is  typical  for
situations in which extra  mixing episodes beyond the fully convective
core during central helium burning are allowed (Straniero et al 2003).
In  particular, the  sharp  variation around  $M_r  \approx 0.56  M_*$
induced by  mechanical overshoot could  be a potential source  of mode
trapping in the core region. The bump in $N$ at $M_r \approx 0.96 M_*$
is other  possible source  of mode trapping,  in this  case associated
with modes trapped in the outer  layers. This feature is caused by the
chemical  transition  of  helium,  carbon and  oxygen  resulting  from
nuclear processing in prior evolutionary stages.

The influence  of the chemical composition gradients  on the pulsation
pattern is clearly shown in panel (a) of Fig.
\ref{deltap}, in which the $\ell= 1$ forward period 
spacing ($\Delta \Pi$) is plotted  in terms of the periods ($\Pi$) for
the same  model analyzed in  Fig. \ref{profile}.  The plot  shows very
rapid variations  in $\Delta \Pi$  everywhere in the  period spectrum,
with  trapping amplitude  (measured as  the interval  in  $\Delta \Pi$
between a period spacing maximum and the adjacent minimum) up to about
6 s  and a trapping cycle  (measured as the interval  in $\Pi$ between
the  period spacing  minima) of  $\approx 70$  s.  The  rather complex
period-spacing diagram shown by Fig. \ref{deltap} is typical of models
characterized by several chemical interfaces.  In order to disentangle
the effect of each chemical  composition gradient on mode trapping, we
follow  the procedure of  Charpinet et  al. (2000). Specifically, 
we minimize  --- although no completely eliminate
--- the effects  of a given  chemical interface simply by  forcing the
Ledoux term $B$ to be zero in the specific region of the star in which
such interface  is located. In  this way, the resulting  mode trapping
will be only due  to the remainder chemical interfaces.  Specifically,
we have  recomputed the entire  $g$-mode period spectrum  assuming (1)
$B=  0$ at  the  region  of the  O/C/He  chemical interface  ($M_r/M_*
\approx 0.96$), and (2) $B= 0$ at the region of O/C chemical interface
($M_r/M_* \approx 0.4-0.6$) (see inset of Fig.  \ref{profile}).  The 
results are shown in panels  (b) and (c) of Fig.   \ref{deltap}, 
respectively.  By
comparing  the different  cases illustrated,  an  important conclusion
emerges  from this  figure:  the chemical  transition  region at  $M_r
\approx 0.96  M_*$ is responsible for the  non-uniformities in $\Delta
\Pi$ only for $\Pi
\lesssim 500 $ s (panel c), whereas the chemical composition gradients
in the core region ($M_r \approx 0.4-0.6 M_*$) cause the mode-trapping
structure in the rest of the period spectrum (panel b).

From the  above discussion,  it is clear  that {\it  the mode-trapping
features predicted by full evolutionary PG1159 models are dominated by
the core chemical structure left by prior overshoot episodes.  This is
particularly  true  for  the  range  of periods  observed  in  GW  Vir
stars}. On its hand, the more external chemical transition has a minor
influence, except  in the  regime of short  periods.  This  finding is
clearly at odds  with previous results reported by  Kawaler \& Bradley
(1994).   Indeed,  these authors  have  found  that the  mode-trapping
properties of their PG1159 models are fixed mainly by the outer O/C/He
transition region, to such a degree that they have been able to employ
mode-trapping  signatures as  a sensitive  locator of  this transition
region.  We  note that our  PG1159 stellar models  differ considerably
from  those  employed  by  Kawaler  \&  Bradley  (1994),  particularly
concerning  the details of  the treatment  of the  evolutionary stages
that lead to the formation of PG1159 stars.  Of particular interest is
the presence of a much  less pronounced chemical transition in the C/O
core of  the Kawaler  \& Bradley (1994)  models, as compared  with the
rather  abrupt overshoot-induced  chemical gradients  at  $M_r \approx
0.4-0.6 M_*$ in our full  PG1159 evolutionary models.  In addition, we
note that  our evolutionary treatment predict  thick helium envelopes.
Thus, trapping structure predicted by  our models is due mostly to the
core chemical  gradients.  If we  artificially minimize the  effect of
these  gradients we  immediately  recover the  results  of Kawaler  \&
Bradley (1994) for the case of thick helium envelopes. 

In addition to the issue of core overshooting, other relevant 
point regarding the core chemical stratification of our models are 
the adopted reaction rates during the central helium burning phase. 
Following a suggestion of an anonymous referee, we have explored the 
possibility of a smaller rate of the $^{12}$C$(\alpha,\gamma)^{16}$O 
nuclear reaction (see Kunz et al. 2002). We found 
that with a lower rate of that reaction, the resulting central 
abundances of oxygen and carbon are quite similar, partially 
smoothing the chemical steps  in the core. However, appreciable 
structure remains that is still able to give clear pulsational 
signals associated to the occurrence of prior core overshooting.   

\begin{figure}
\centering
\includegraphics[clip,width=250pt]{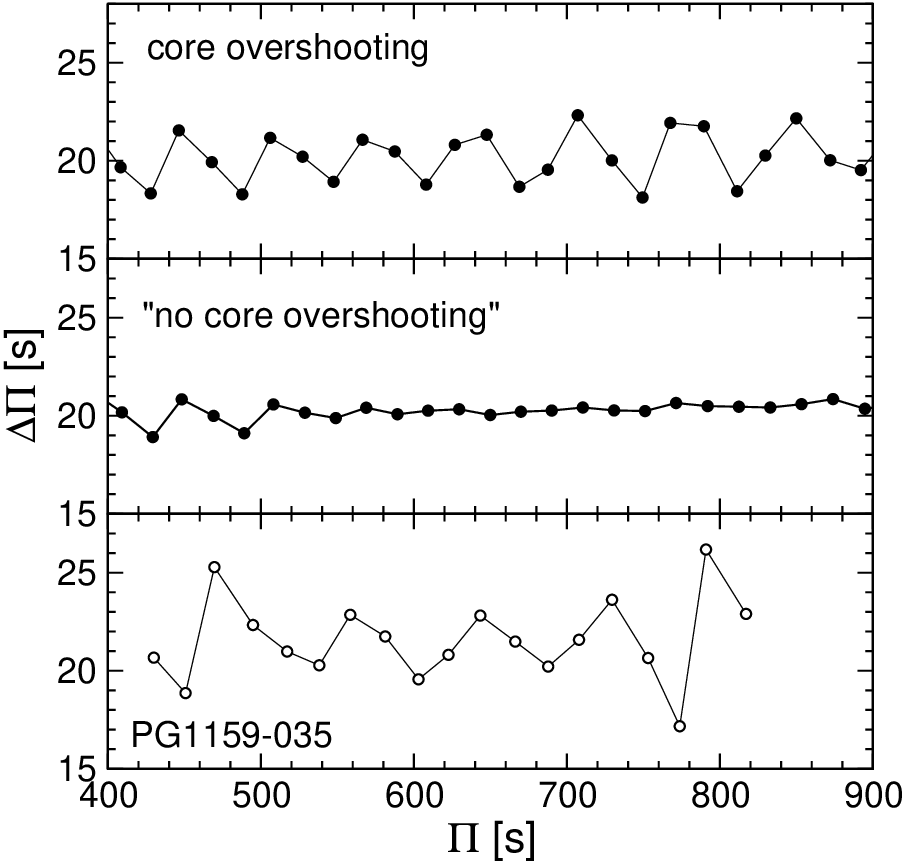}
\caption{The forward period spacing $\Delta \Pi$ vs period 
$\Pi$ for dipole ($\ell= 1$) modes for the same model analyzed in Fig.
\ref{deltap}. Upper panel corresponds to 
pulsational  computations in  which the  Ledoux term  $B$  is computed
self-consistently (core overshooting), whereas middle panel correspond
to the  case in which $B=  0$ at $M_r/M_* \approx  0.4-0.6$ (``no core
overshooting'').   The  observed period  spacings  for PG1159-035  are
plotted in lower panel.  See text for details.}
\label{deltapg2}
\end{figure}

In Fig. \ref{deltapg2} we  compare the predictions of our calculations
with the pulsational  spectrum of PG1159-035, the prototype  of GW Vir
stars. We do not intend  to perform here a detailed asteroseismological
fit  to  this  star. Instead,  we  want  to  show that  the  structure
associated  with the observed  period spacing  of PG1159-035  could be
reflecting the  presence of  chemical gradients in  its inner  core, a
fact that is  borne out by Fig. \ref{deltapg2}. In  fact, we find that
according to the  predictions of the theory of  post-AGB evolution for
the  chemical  stratification  for  the pulsating  PG1159  stars,  the
occurrence of core overshoot episodes during central helium burning is
required to be consistent with seismological demands of these stars.

We  judge that  the  period-spacing distribution  exhibited by  PG1159
stars bears the signature of core overshoot episodes and that improved
observations  would eventually turn  these pulsating  pre-white dwarfs
into a powerful tool for shedding new lights on the mixing and central
burning  processes   occurred  in  the  white   dwarf  progenitors  of
intermediate  masses. However,  it is  not unconceivable  that stellar
winds  during  the  PG1159  stage  could  reduce  the  helium  content
considerably, with  the consequent  result that the  trapping features
induced  by core  overshooting  could not  be  disentangled from  that
caused by the O/C/He transition zone.


\begin{acknowledgements}

This   research  was   partially   supported  by   the  Instituto   de
Astrof\'{\i}sica La Plata. L.G.A.  acknowledges the Spanish MCYT for a
Ram\'on y Cajal Fellowship.

\end{acknowledgements}

\end{document}